\pdfoutput=1 
\documentclass{JINST}
\usepackage{cite}

\title{Physics of Fully Depleted CCDs}

\author{S. E. Holland\thanks{Corresponding author.},~
C.J. Bebek, W.F. Kolbe, 
and J.S. Lee\\
\llap{$^a$}Lawrence Berkeley National Laboratory\\
  1 Cyclotron Road, Berkeley, CA, 94720, USA\\
E-mail: \email{seholland@lbl.gov}}

\abstract{In this work we present simple, physics-based models
for two effects that have been noted in the fully depleted CCDs
that are presently used in the Dark Energy Survey Camera. The
first effect is the observation that the point-spread function
increases slightly with the signal level.  This is explained
by considering the effect on charge-carrier diffusion due to the 
reduction in the magnitude of the channel potential as collected 
signal charge acts to partially neutralize the fixed charge in the 
depleted channel. The resulting reduced voltage drop across the
carrier drift region decreases the vertical electric field and
increases the carrier transit time.  The second effect is the observation 
of low-level, concentric ring patterns seen in uniformly illuminated images.  
This effect is shown to be most likely due to lateral deflection of charge 
during the transit of the photo-generated carriers to the potential wells as 
a result of lateral electric fields.  The lateral fields are a result of 
space charge in the fully depleted substrates arising from resistivity 
variations inherent to the growth of the high-resistivity silicon used to 
fabricate the CCDs.}


\begin{document}

\section{Introduction}\label{sec:intro}

Fully depleted charge-coupled devices (CCDs) are presently in use in three 
major astronomical imaging cameras: the PAN-Starrs PS1 
camera~\cite{panstarrs}, the HyperSuprime-Cam camera~\cite{hsuprime}, and the 
Dark Energy Survey Camera (DECam)~\cite{des}.  The thicknesses of the CCDs are 
75~\cite{panstarrs2}, 200~\cite{hsuprime2} and 250~$\mu$m~\cite{des2}, 
respectively.  The primary advantages of the thick, fully depleted devices 
compared to thinned, conventional scientific CCDs are the enhanced quantum 
efficiency and reduced fringing at long wavelengths~\cite{groom1999}.

The purpose of this work is to present first order, physics-based models to 
explain some of the low-level phenomena that have been observed during 
telescope operation with these thick, fully depleted CCDs.  The effects of 
interest in this work are the dependence of the CCD point-spread function 
(PSF) on signal level, and the appearance of fixed-pattern, low-level 
variations in the CCD images seen during uniform illumination of the devices.  
For the former we present a first order theoretical model for the effect, while 
for the latter we include experimental results to assist in the understanding of 
the basic mechanisms involved.

The CCDs described in this work are fabricated on high-resistivity silicon
substrates that are manufactured using the float-zone refining
technique~\cite{Ammon}.  The resistivity is greater than
4000~$\Omega$-cm for the n-type substrates considered here, and the CCDs are
operated with a substrate-bias voltage that fully depletes the high-resistivity
substrate~\cite{Holland1}.  As discussed in the remainder of the paper,
the low-level effects that arise in these thick, fully depleted CCDs depend 
on the vertical electric fields in the device, and we show that the effects can
be reduced by operating the CCDs with increased vertical fields. 

\section{Effect of signal charge on the point spread function}\label{sec:PSF}

\begin{figure}[tbp] 
\centering
\includegraphics[width=0.9\textwidth]{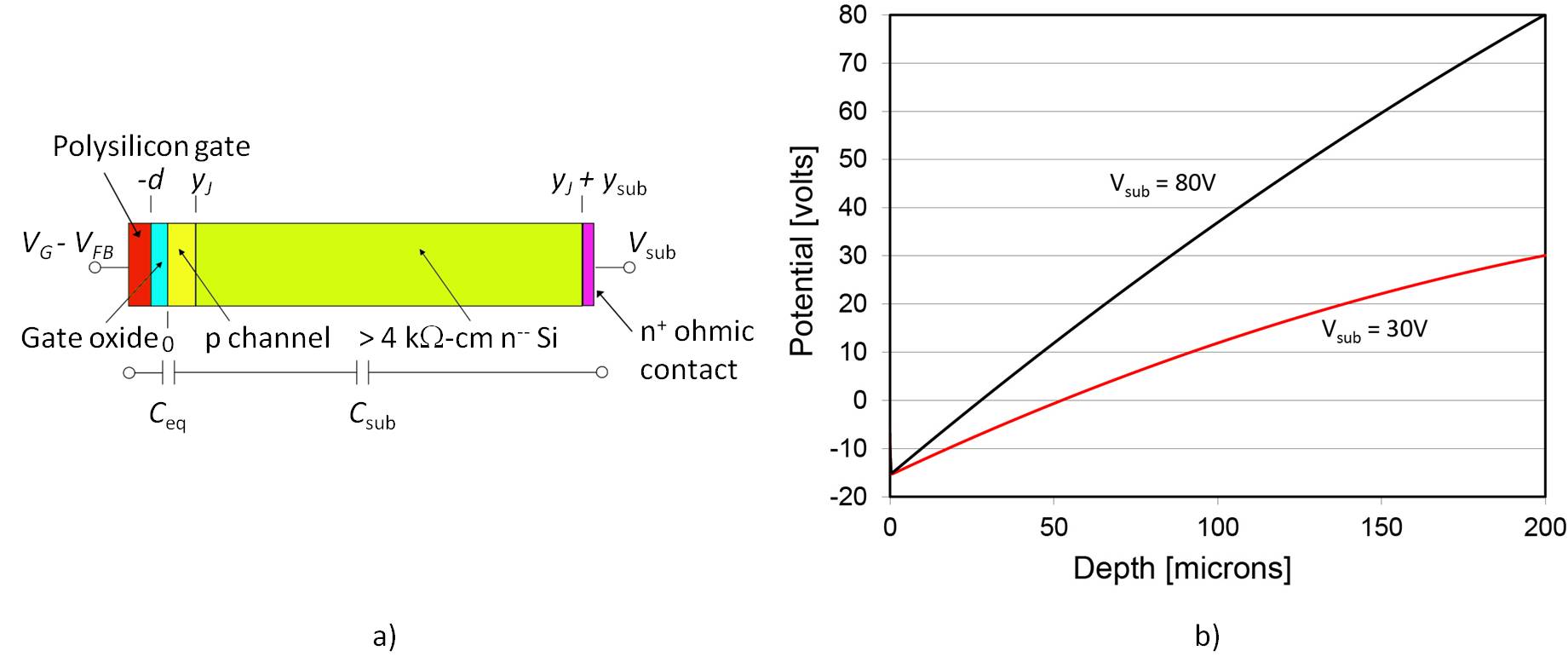}
\caption{a) Simplified cross-sectional diagram for a fully depleted CCD.  b) Simulated 
electrostatic potential for a 200~$\mu$m thick CCD at substrate-bias voltages of 
30 and 80~V.}
\label{fig:cross}
\end{figure}

We have previously described simple models for the PSF of fully depleted 
CCDs~\cite{Holland_Bruges,groom_eso,Holland_2003,Karcher_et_al,Fairfield_et_al}.
The model is based on the lateral diffusion of the photo-generated holes as
they are drifted by the vertical electric field in the fully depleted substrate.
We briefly review this model and then discuss enhancements to take into account
the effect of signal charge.  The PSF in the over-depleted 
case is given by

\begin{equation}
\label{eq:sigma}
PSF_{od} = \sqrt{2 D \, t_{\rm{tr}}} = \sqrt{2 {kT \over q} {\epsilon_{\rm{Si}} \over \rho_n} \ln {E_{\rm{y_J}} \over {E_{\rm{y_{sub}}} } } }
\end{equation}

\noindent where $D$ and $t_{\rm{tr}}$ are the diffusion coefficient and
transit time for holes, respectively, $k$ is Boltzmann's constant, $T$ is the
temperature in kelvin, $q$ is the electron charge, $\epsilon_{\rm{Si}}$ is the
permittivity of silicon, and $\rho_n=qN_D$ is the volume charge density in
the fully depleted substrate where $N_D$ is the substrate doping density.
Referring to Fig.~\ref{fig:cross}~a), ${E_{\rm{y_J}}}$ is the electric field
at the p-channel, n-substrate junction located at $y_J$, and ${E_{\rm{y_{sub}}} }$
is the electric field at the back-side contact of the CCD.

The electric field in the fully depleted substrate from a 
one-dimensional (1-D) solution of the Poisson equation assuming uniform doping 
density in the channel and substrate is given by~\cite{Holland_2003} 

\begin{equation}
E(y) = {{q N_D} \over {\epsilon_{\rm{Si}}}} {(y - y_J)} + E_J \ \ \ \ \ y_J < y < (y_J + y_{\rm{sub}}) 
\label{eq:field_substrate} 
\end{equation}

\noindent where

\begin{equation}
{E_J} \equiv - {{{dV} \over {dy}} (y_J)} = - \bigg( {{V_{\rm{sub}} - V_J} \over {y_{\rm{sub}}}}
+ {1 \over 2 } {{q N_D} \over \epsilon_{\rm{Si}}} {y_{\rm{sub}}} \bigg) 
\label{eq:field_junction_defined} 
\end{equation}

\noindent $V_{\rm{sub}}$ is the applied substrate-bias voltage, and $V_J$ is
the electrostatic potential at the p-channel, n-substrate 
junction.  It was shown in reference~\cite{Holland_2003} that $V_J$ is 
approximately equal to the value of the CCD potential at the potential 
minimum $V_m$, and that $V_J$ is approximately independent of 
$V_{\rm{sub}}$ when the substrate doping is much less than the channel 
doping and the substrate thickness is much greater than the sum of the 
channel and gate-insulator thickness.  A simple capacitor-voltage divider 
equivalent circuit makes this clear.  $C_{\rm{sub}}$ in Fig.~\ref{fig:cross}~a) 
is much less than the equivalent series capacitance $C_{\rm{eq}}$ of the 
channel and gate dielectric, and as a result the channel potential is a 
weak function of the substrate voltage.

Figure~\ref{fig:cross}~b) shows the simulated electrostatic potential 
versus depth for a 200~$\mu$m thick CCD at substrate-bias voltages of 30 and 
80~V.  The simulation cross-section was generated from a two-dimensional simulator 
that takes into account the steps used in fabricating the actual devices.  The 
simulation methods are described in more detail below.  A key point is that the
total voltage drop across the depleted region 
$V_{\rm{sub}}-V_J \approx V_{\rm{sub}}-V_m$ determines
the electric field and hence the PSF.  In practice when fitting the model 
to experimental data we have found that the dependence of hole mobility on 
electric field must be taken into account in order to accurately fit data at 
high substrate-bias voltages~\cite{Fairfield_et_al}.

The model presented above assumes that $V_J$ is constant and independent
of $V_{\rm{sub}}$.  While the latter is a good approximation, the former
is not valid when charge is stored in the CCD potential wells.  Early
numerical~\cite{kent_1973} and analytical~\cite{el-sissi_cobbold_1974} 
solutions to the 1-D Poisson equation that include the signal charge
in the CCD potential wells showed that the effect of signal charge is
to decrease in magnitude $V_m$.  We postulate that this effect 
is the cause of the observed dependence of the PSF on the signal level 
that has been observed during the operation of the Dark Energy Survey 
Camera.  In order to study this in more detail, we incorporate the effect 
of the signal charge on $V_m$ in the simple PSF model as described below.

The 1-D models described in references~\cite{kent_1973,el-sissi_cobbold_1974} 
predict an approximately linear dependence on the signal-charge level for
the potential at the Si-SiO$_2$ interface and at the location of the potential
minimum, respectively.  An analytic equation for this relationship was given
in reference~\cite{el-sissi_cobbold_1974} with $V_m$ dependent on the channel 
doping and thickness, the gate-insulator thickness, the substrate doping, 
the applied gate voltage, and the signal-charge level.

\begin{figure}[tbp] 
\centering
\includegraphics[width=0.9\textwidth]{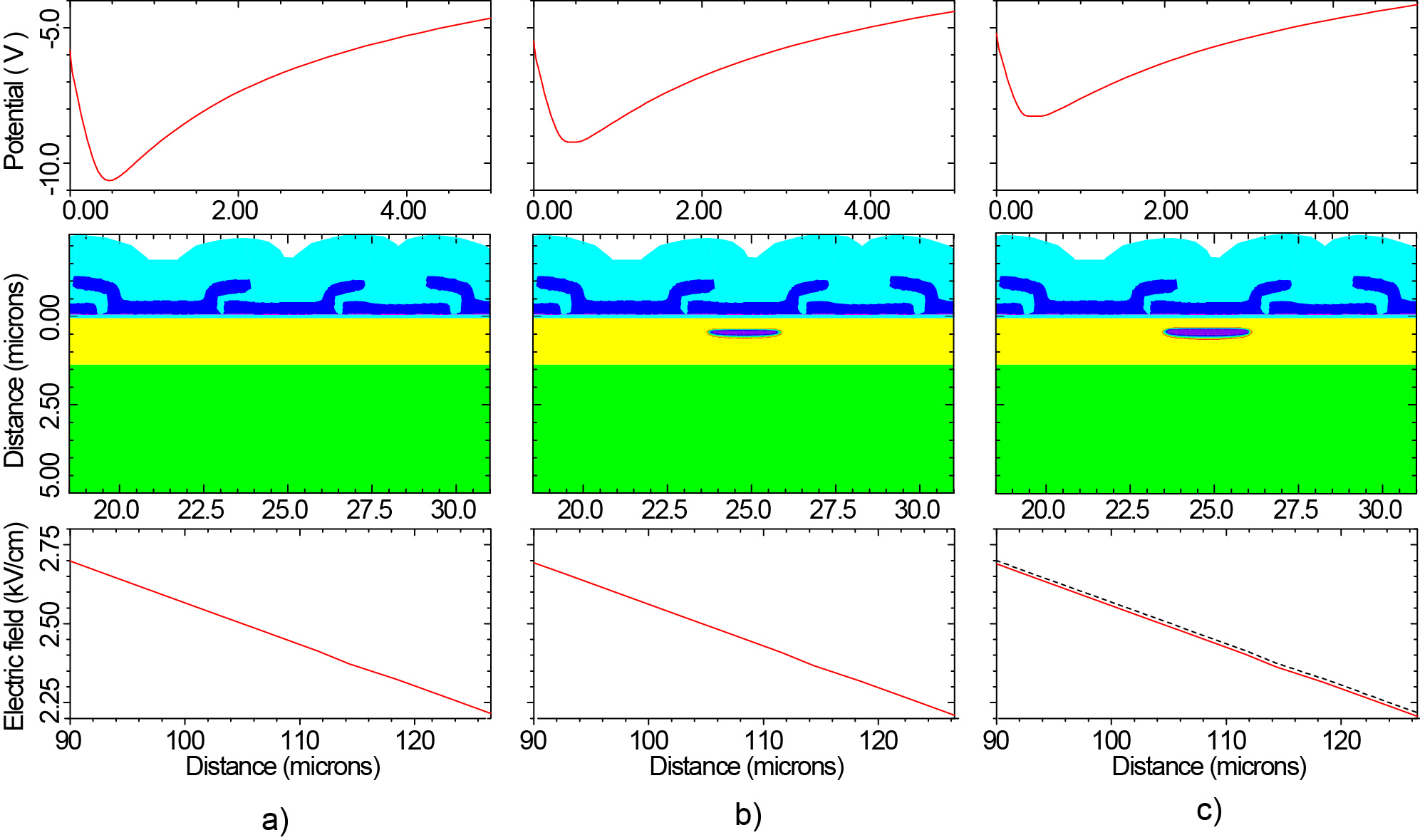}
\caption{Two-dimensional simulation results demonstrating the change in
the vertical electric field (lower plots) and the electrostatic potential
near the surface (upper plots) as charge is added to the CCD potential well.
The charge levels from left to right are 0, 4120 and 8090 holes/$\mu$m width. 
The center panel shows the simulated device cross-section, and the charge
packet of holes is represented as a logarithmic contour plot.  The yellow
region is the p-type channel, and the green region corresponds to the 
high-resistivity, n-type substrate.  The dark/light blue features are the
polysilicon and SiO$_2$ regions, respectively.  The potential is plotted versus
depth along the center line of the collecting phase.  As in the simplified 1-D model,
the origin for the potential plots is the Si-SiO$_2$ interface, and the 
vertical electric field is plotted over the depth range of 90 to 
130~$\mu$m from the Si-SiO$_2$ interface.  The dashed line in the electric field
plot in c) is taken from a), i.e. the case for no signal charge. }
\label{fig:psf_sims}
\end{figure}

We have studied the effect of signal charge on the channel potential with the
aid of 2-D simulation.  Figure~\ref{fig:psf_sims} shows results from the
simulations.  The process simulator Synopsys TSUPREM-4  was used to generate 
a realistic cross-section.  Photolithography, etching, ion implantation, 
oxidation, thin-film deposition, and diffusion steps as specified
in the actual fabrication sequence were included in the simulation.  This
cross-section including accurate doping profiles and geometries
was then input into the 2-D device simulator Synopsys Medici that solves
the Poisson and semiconductor continuity equations in a self-consistent fashion.
In addition, the 2-D device simulator can model photo-generation of charge.

The middle panel of Fig.~\ref{fig:psf_sims} shows a portion of the cross-section
that was generated and used in the 2-D device simulation.  The full cross-section
includes 3 pixels at a 10.5~$\mu$m pitch.  Three layers of polysilicon are
used in the 3-phase CCDs under study here.  The cross-sectional plot shows 
the spatial distribution of the signal charge represented as a logarithmic 
contour plot.  

The upper portion of Fig.~\ref{fig:psf_sims} shows the effect 
of adding signal charge on the electrostatic potential that is shown plotted 
versus depth along the center line of the collecting phase.  The origin of 
the plot corresponds to the Si-SiO$_2$ interface.  The effect of the signal 
charge is to electrically neutralize a portion of the depleted channel, 
i.e.  the holes collected in the channel neutralize an equal amount of 
ionized acceptor atoms resulting in a charge-neutral region where the 
field is zero.  As charge is added to the pixel $V_m$ rises.  This in 
turn reduces the voltage drop $V_{\rm{sub}}-V_J \approx V_{\rm{sub}}-V_m$ 
across the drift region which in turn reduces the vertical electric field 
as shown in the lower portion of Fig.~\ref{fig:psf_sims}.  The reduction 
in vertical electric field causes the hole-transit time to increase resulting 
in an increase in the PSF.  This is the essence of the model that we
are presenting here.

\begin{figure}[tbp] 
\centering
\includegraphics[width=1.0\textwidth]{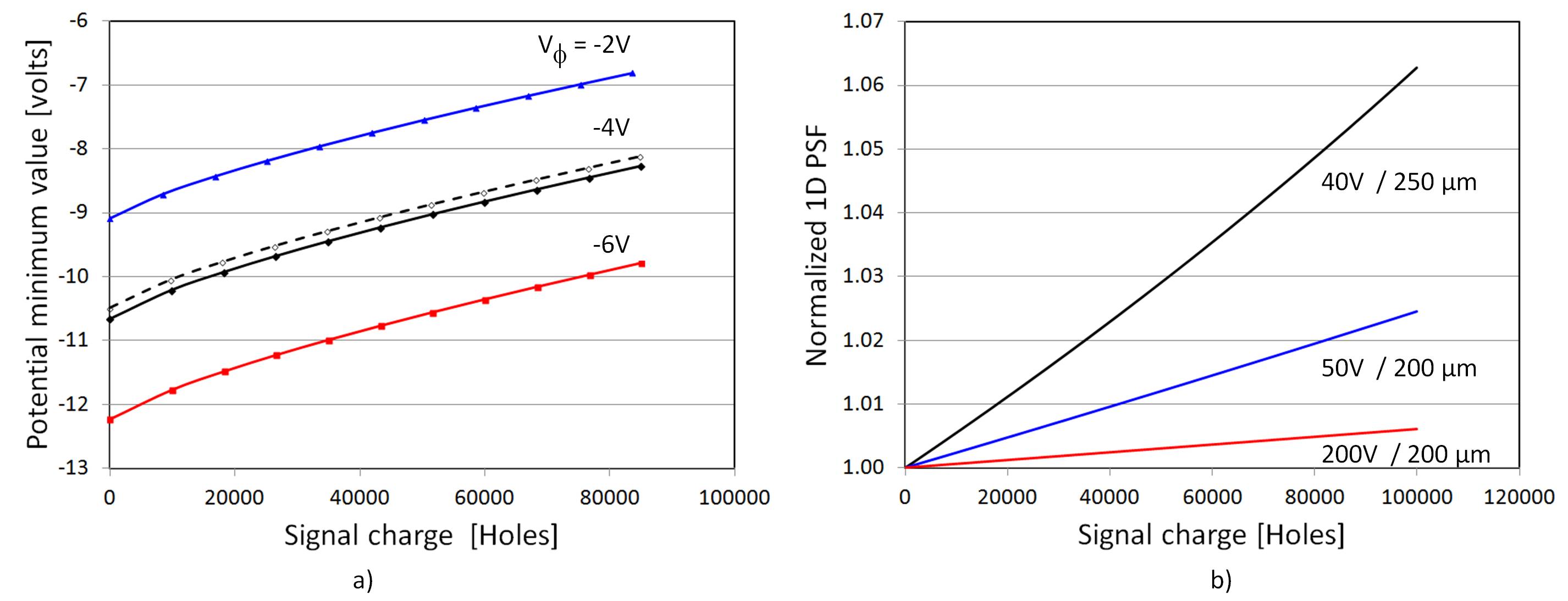}
\caption { a) Simulated potential-minimum value $V_m$ versus signal-charge level for
different values of substrate and clock voltages.  The solid and dashed lines 
are for $V_{\rm{sub}}$= 50 and 100~V, respectively. b) Calculated normalized PSF 
versus signal-charge level for various combinations of substrate-bias voltage
and CCD thickness.  The details of the calculation are described
in the text.  For both plots the simulated charge per unit width was scaled by a width 
factor of 10.5~$\mu$m.}
\label{fig:vm_psf_vs_qs}
\end{figure}

To explore the proposed model in more detail, Fig.~\ref{fig:vm_psf_vs_qs}~a) 
shows the simulated $V_m$ versus signal charge.  Consistent with the earlier 1-D 
models the simulated $V_m$ is an approximately linear function of the signal 
charge~\cite{kent_1973,el-sissi_cobbold_1974}.  Simulations were done at
substrate-bias voltages of 50 and 100~V, and as seen in 
Fig.~\ref{fig:vm_psf_vs_qs}~a) the substrate-bias voltage has a small effect
on $V_m$ even when signal charge is stored in the pixel.


Figure~\ref{fig:vm_psf_vs_qs}~b) shows the calculated PSF as a function of
signal charge.  The plot was generated by calculating the value of $V_m$
for a given signal-charge level $q_s$ from a linear fit to simulations
of the type shown in Fig.~\ref{fig:vm_psf_vs_qs}~a).  This value of 
$V_m=f(q_s)$ was then substituted into Eqs.~\ref{eq:sigma}
through ~\ref{eq:field_junction_defined} to generate the calculated PSF versus 
$q_s$ plot shown in Fig.~\ref{fig:vm_psf_vs_qs}~b).

The linearity of the predicted PSF versus $q_s$ plots shown in 
Fig.~\ref{fig:vm_psf_vs_qs}~b) improves as the substrate-bias voltage is increased,
and the slope decreases.  At large $V_{\rm{sub}}$ the space charge due to ionized 
dopants in the fully depleted substrate can be ignored, and Eq.~\ref{eq:sigma} 
becomes~\cite{Holland_Bruges,Holland_2003}

\begin{equation}
\label{eq:sigma_a}
PSF_{od} \approx  \sqrt{2 {kT \over q} { {y_{\rm{sub}}}^2 \over {V_{\rm{sub}}-V_J}}}
= \sqrt{2 {kT \over q} { {y_{\rm{sub}}}^2 \over {V_{\rm{sub}}}}
{ \left( 1 - {V_J \over V_{\rm{sub}}} \right)^{-1} } }
\end{equation}

\noindent The last term in the above equation is of the form 
$1/\left(1-x\right)$ which is approximately equal to $\left(1+x\right)$ for
$V_J \ll V_{\rm{sub}}$.  The square root of $\left(1+x\right)$ is approximately
$\left(1+ x/2\right)$ for small $x$.  With these approximations Eq.~\ref{eq:sigma_a}
becomes

\begin{equation}
\label{eq:sigma_a2}
PSF_{od} \approx \sqrt{2 {kT \over q} { {y_{\rm{sub}}}^2 \over {V_{\rm{sub}}}} }
{ \left( 1 + {1 \over 2} {V_J \over V_{\rm{sub}}} \right) }
\approx \sqrt{2 {kT \over q} { {y_{\rm{sub}}}^2 \over {V_{\rm{sub}}}} }
{ \left( 1 + {1 \over 2} {V_m \over V_{\rm{sub}}} \right) } 
\ \ \ \ \ V_J \ll V_{\rm{sub}}
\end{equation}

\noindent which yields a linear dependence on $V_m$.
Since $V_m$ is approximately a linear function of the signal charge $q_s$,
it follows that the PSF is approximately a linear function of $q_s$ as well 
when the simplifications described above are valid. The predicted slope of the 
PSF versus $q_s$ plot is then

\begin{equation}
\label{eq:slope_psf}
{{d \left( PSF_{od} \right) } \over {d q_s}} \approx  \beta \, { {y_{\rm{sub}}} \over {V_{\rm{sub}}} }
\sqrt{{ 1 \over 2} {kT \over { q {V_{\rm{sub}}} } } }
\end{equation}

\noindent where $\beta$ is the slope of the $V_m$ versus $q_s$ relationship that
depends on the channel doping and other parameters as described earlier.  

The model presented above is clearly a simplified view of the situation,
and regrettably we do not have experimental data at the present
time.  Nonetheless, we present the model in order to hopefully 
elucidate the basic physics of the problem.  The model does explain qualitatively 
the effect of increasing signal charge leading to increased PSF, and does 
predict that these effects will be a function of the CCD thickness and applied 
substrate voltage.  A more accurate model would take into account the cumulative
degradation of the PSF as signal charge is added.

\section{Low-level effects in uniform illumination}\label{sec:rings_sec}

\begin{figure}[tbp] 
\centering
\includegraphics[width=1.0\textwidth]{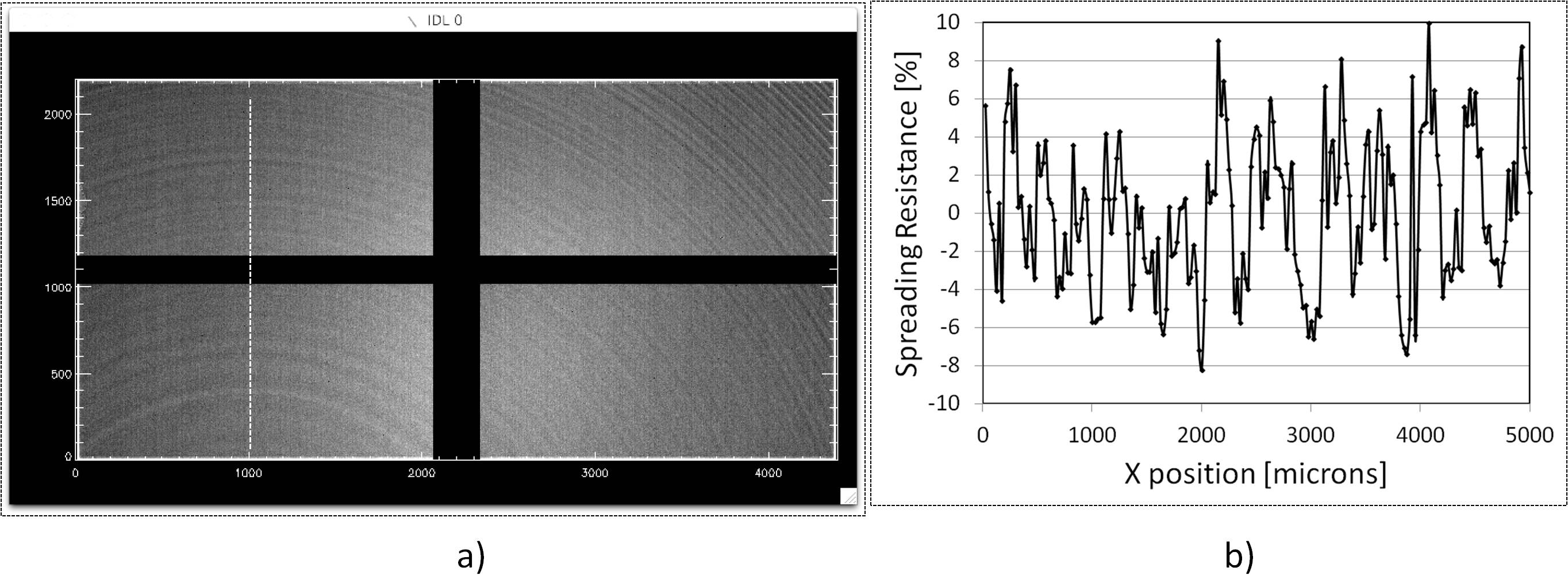}
\caption{a) Processed image taken with a 4114~$\times$~2040, 250~$\mu$m thick, 
15~$\mu$m pixel, fully depleted CCD.   The substrate-bias voltage was 50~V, and the 
operating temperature was -140$~^{\rm{o}}$C.  The CCD was uniformly illuminated with
500~nm light, and 100 images were median filtered to generate the picture shown.
The mean light level was about 30,000 photons/pixel.  The vertical dotted line at 
column 1000 denotes the radial direction used in the analysis described in the 
text.  b) The percent variation in spreading resistance measured on a 
300~$\Omega$-cm, n-type, float-zone refined sample. The step size
for the 2-point probe measurement was 25~$\mu$m.}
\label{fig:rings}
\end{figure}

The CCDs described in this work have been shown to exhibit low-level variations
in the signal level when illuminated uniformly with light.  Figure~\ref{fig:rings}~a) 
shows an image resulting from the median-filtering of 100 exposures taken with 
a 4114~$\times$~2040, 250~$\mu$m thick, 15~$\mu$m pixel, fully depleted CCD that 
was uniformly illuminated with 500~nm light.  A fixed-pattern feature consisting of 
concentric rings is visible in the image that has been displayed with enhanced 
contrast.  As shown later, the variations in the image are on the order of a percent.

Similar fixed patterns have been reported for n-channel, fully depleted 
CCDs~\cite{kotov_spie} and other silicon-based imagers including partially depleted, 
back-illuminated silicon vidicons~\cite{crowell_and_labuda,yoshikawa_1973}
and front-illuminated commercial CCDs~\cite{jastrzebski_1980,hiroshima_1984}, Charge
Priming Devices~\cite{senda_1985}, and CMOS image sensors~\cite{Rantzer_2002}.
Concentric patterns that represent the error in position location for charged
particles in silicon drift detectors have also been 
reported~\cite{Nouais_2001,Nouais_2003,Crescio_2005}.
In all cases the underlying mechanism for the fixed pattern is variations 
in the resistivity in the bulk silicon substrates that are inherent to the crystal 
growth processes.  Dopant impurities are incorporated into the silicon during crystal 
growth from the molten phase according to the impurity segregation (distribution) 
coefficient defined as the ratio of impurity solubility in the solid and liquid 
phases of the silicon~\cite{Ammon,Burton_1953}.  Growth-rate fluctuations due to
thermal asymmetries in the melt and temporal variations in temperature that
are affected by the crystal rotation and pull rates result in the resistivity 
striations~\cite{morizane_1967,abe_1973}.  

The concentricity of the rings comes about from the slicing into wafers of the 
bulk ingots transverse to the growth direction where the growth surface, i.e the 
crystal-melt interface is approximately spherical in 
shape~\cite{morizane_1967,abe_1973}.  As a result it is typically observed that 
the spatial frequency of the striations increases towards the outer portion
of the wafer~\cite{hiroshima_1984,senda_1985,kopanski_1992}.  This effect 
can be seen in the upper right corner of the image shown in 
Fig.~\ref{fig:rings}~a).  Also, the resistivity varies in the vertical dimension
as a consequence of the shape of the growth surface.  The resistivity striations
are typically characterized by spreading-resistance 
measurements~\cite{burtscher_1974,abe_1973,kopanski_1992,muhlbauer_1975}.
Fig.~\ref{fig:rings}~b) shows measurements taken on a 
300~$\Omega$-cm, n-type, float-zone refined sample that is believed to be 
qualitatively similar to the silicon used in this work~\cite{leif}.


The concentric fixed patterns in the imagers mentioned previously have been attributed
to lateral electric fields that displace the charge 
carriers~\cite{yoshikawa_1973,jastrzebski_1980}, recombination
centers and crystal defects that affect the amount of charge 
collected~\cite{jastrzebski_1980,Rantzer_2002}, and resistivity-induced 
variations in the potential barrier to the substrate for the imagers that 
utilize a vertical-overflow drain~\cite{hiroshima_1984,senda_1985}.  Of those effects, 
it is expected that the lateral electric fields will dominate in fully depleted CCDs  
as is the case for the silicon drift 
detectors~\cite{Nouais_2001,Nouais_2003,Crescio_2005}.
The float-zone refined silicon used in this work has very few crystalline defects
and carrier traps, and the fully depleted operation reduces the possibility
of recombination. 

 


In references~\cite{yoshikawa_1973,jastrzebski_1980} the lateral electric field 
in the {\em{undepleted}} silicon was modeled as a built-in field that arises from the
detailed balance of diffusion and drift currents in thermal equilibrium.  The
impurity gradient generates a diffusion current that is balanced by the field
that arises when charge carriers diffuse away from their origin leaving behind
ionized dopant atoms~\cite{muller_kamins}.  The field in that case is given by
$E_{\rm{\,x,\,undepleted}} \approx (kT/q) (1/x) \, ln( N_{\rm{D1}}/N_{\rm{D2}})$ 
where $x$ is the lateral spatial coordinate over which the doping density varies 
from $N_{\rm{D1}}$ to $N_{\rm{D2}}$.  

For the case of a fully depleted device, the effect of the resistivity variation is 
to add a lateral variation to the volume charge density $\rho_n=qN_D$ where
$N_D$ is now a function of $x$.  Due to the complexity of solving the 2-D Poisson
equation analytically, one must resort to numerical methods in order to
accurately model the lateral electric field $E_x$ due to the resistivity 
striations~\cite{cornu_1975,kotov_2006}.  We qualitatively estimate the lateral-charge 
displacement in a fully depleted CCD by first calculating the transverse distance 
traveled by the holes due to an assumed constant lateral field $E_x$ during the 
transit time $t_{tr}$ of the holes from their generation point to the collection 
well, i.e.


\begin{equation}
\label{eq:lateral_x}
v_x = {dx \over dt} = \mu_p E_x 
\end{equation}

\noindent where $v_x$ is the hole velocity in the lateral direction $x$.  
After separation of variables and integrating, Eq.~\ref{eq:lateral_x} 
yields the lateral displacement $x_L$ during the time $t_{tr}$

\begin{equation}
\label{eq:lateral_x2}
x_L = \mu_p \, E_x \, {t_{tr}}
\end{equation}

\noindent The transit time neglecting the space charge in the substrate is

\begin{equation}
\label{eq:ttr}
t_{tr} = { {y_{\rm{sub}} }^2 \over  \mu_p \left( {{V_{\rm{sub}} - V_J} } \right) }
\end{equation}

\noindent The fractional, lateral displacement of the holes is then solved for by
substituting Eq.~\ref{eq:ttr} into Eq.~\ref{eq:lateral_x2} yielding

\begin{equation}
\label{eq:f_xl}
{ x_L \over y_{\rm{sub}} } =  { {E_x \: y_{\rm{sub}} } \over {{V_{\rm{sub}} - V_J} } } 
\approx { E_x \over E_y }
\end{equation}

\noindent A similar argument was presented in reference~\cite{jastrzebski_1980} for the
case of the built-in field that applies in undepleted regions.

This simple model predicts that the amplitude of the deviations of the light level from
the mean value are independent of temperature and can be reduced by increasing the 
substrate-bias voltage.  The residual-light level should also be wavelength 
dependent, with shorter wavelength light showing the largest effect since the 
photo-generated holes travel the longest distance in this case and will be subject 
to the lateral fields for a longer period of time.  In the following we explore 
the validity of Eq.~\ref{eq:f_xl} by comparing the predictions to experimental results. 

In order to improve the statistics of the measurements, median-filtered images generated
from 100 individual exposures were used for the analysis.  The images were then
converted to polar coordinates to transform the circular patterns into linear ones that
were then averaged over 500 pixels for each data point shown in the following.  The
conversion to polar coordinates changes the row, column pixel coordinates to 
$(r, \theta)$, and the averaging is then done over a range of angles 
centered at $\theta$~=~$\pi/2$ over the radial distance $r$ for the data presented below.  
The angle $\pi/2$ corresponds to the dotted line at about column 1000 in 
Fig.~\ref{fig:rings}~a).

\begin{figure}[tbp] 
\centering
\includegraphics[width=1.0\textwidth]{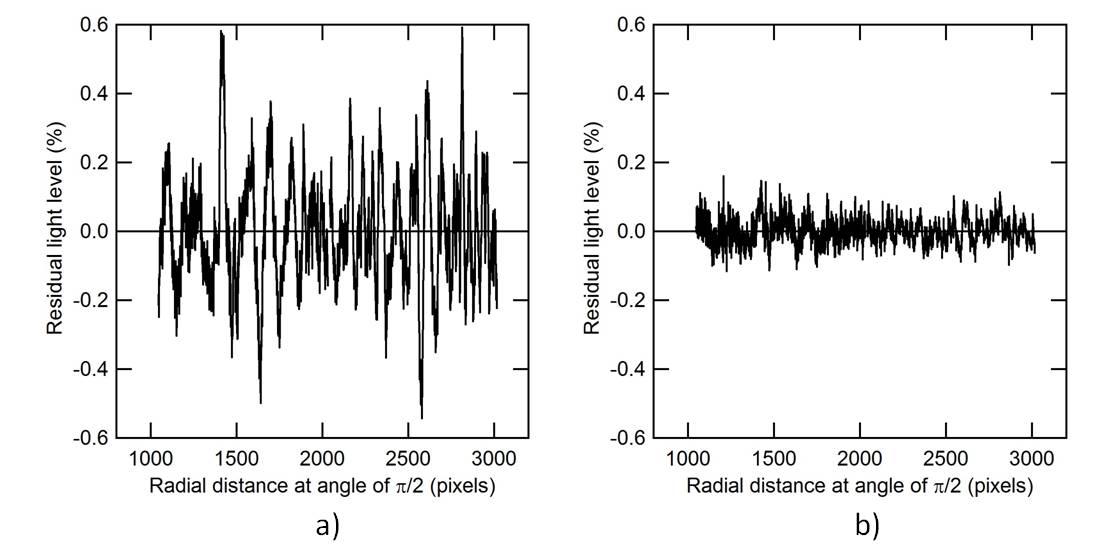}
\caption{The residual-light level in percent measured at two substrate-bias voltages on a
4114~$\times$~2040, 250~$\mu$m thick, 15~$\mu$m pixel CCD.  The operating temperature was
-140$~^{\rm{o}}$C. a) $V_{\rm{sub}}$=40~V.  b) $V_{\rm{sub}}$=150~V.}
\label{fig:rings_vsub}
\end{figure}

\begin{figure}[tbp] 
\centering
\includegraphics[width=1.0\textwidth]{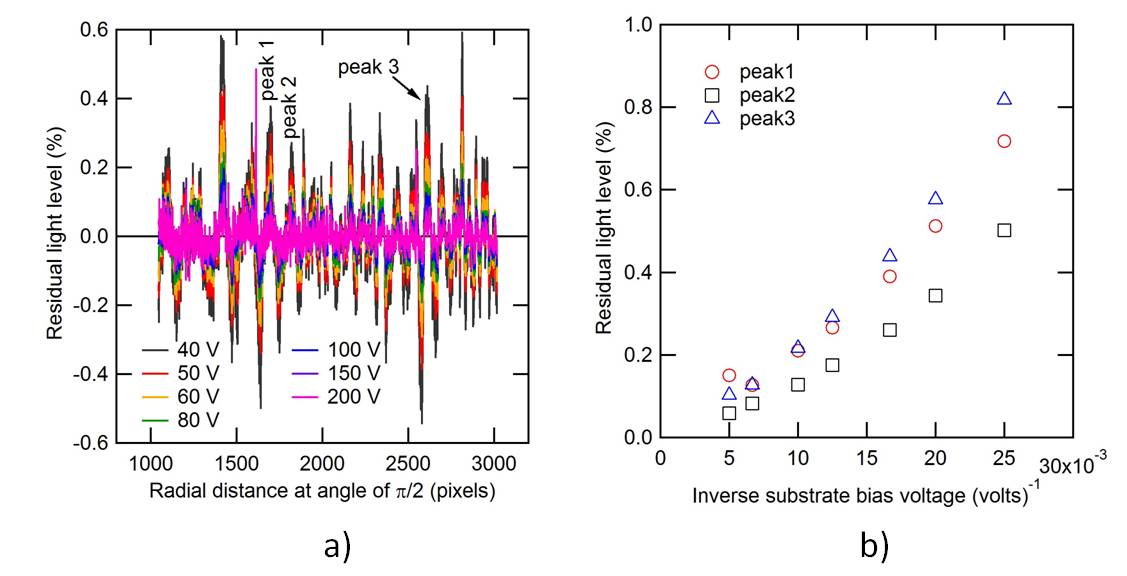}
\caption{a) The residual-light level in percent measured at different substrate-bias
voltage levels on a 4114~$\times$~2040, 250~$\mu$m thick, 15~$\mu$m pixel CCD.
b) The residual-light level peak-to-peak amplitude versus inverse substrate-bias voltage
calculated for peaks 1 to 3 in a).}
\label{fig:1over_vsub}
\end{figure}

Figure~\ref{fig:rings_vsub} shows the residual-light level in percent for 
substrate-bias voltages of 40 and 150~V.  The operating temperature was 
-140$~^{\rm{o}}$C.  We have also performed measurements at -100~$^{\rm{o}}$C and 
do not see a significant temperature dependence.  The residual-light level was 
calculated by first subtracting a smooth curve from the raw data.  This was 
necessary in order to account for small non-uniformities in the light intensity 
that was used to illuminate the CCD.  A quadratic, least-squares fit was used 
to generate the smooth curve.  The residual value in analog-to-digital units  
was then divided by the average light level calculated from the quadratic-fit 
curve to generate the data shown in Fig.~\ref{fig:rings_vsub}. As predicted by 
Eq.~\ref{eq:f_xl} the amplitude of the residual-light level is reduced as 
$V_{\rm{sub}}$ is increased due to the reduction in transit time at higher 
$E_y$ that in turn reduces the time that the holes are exposed to the lateral 
fields due to the resistivity variations.  It must be noted that the CCDs 
described in this work were specifically designed to operate reliably at 
high substrate-bias voltages~\cite{holland_2006,holland_2009}.

Eq.~\ref{eq:f_xl} predicts that the fractional charge displacement should decrease
as 1/$(V_{\rm{sub}}-V_J)$.  Figure~\ref{fig:1over_vsub}~a) shows the measured 
residual-light level in percent showing three regions where the peak-to-peak amplitudes 
were extracted at each substrate-bias voltage. A binomial smoothing algorithm was used to 
find the minimum and maximum values for each region.  Figure~\ref{fig:1over_vsub}~b) 
shows a plot of the extracted amplitude of the residual-light level in percent as a 
function of inverse substrate-bias voltage over the $V_{\rm{sub}}$ range of 40 to 
200~V.  The data trend in qualitative agreement with the simple model presented above. 
The current model neglects the spatially varying transverse diffusion due to the 
resistivity striations and the effect of the vertical electric field on the hole 
mobility. Also, the assumption of a constant $E_x$ is an 
oversimplification~\cite{kotov_2006}.  These effects will be incorporated in future 
studies.  Nonetheless the simple model does point towards two means to reduce the 
effect: operation at high vertical electric fields, and the use of the highest 
resistivity silicon that is practical in order to reduce the magnitude of the
lateral fields.

\section{Summary}\label{sec:summary}

We have presented a simple, physics-based model to explain the dependence of the 
point-spread function on light level in fully depleted CCDs.  Given that this work 
is part of the proceedings of a workshop, it is not intended that this model be 
considered a rigorous solution to the problem but rather a starting point for a 
more accurate model.  We have also studied the low-level variations in signal level 
seen during uniform illumination of fully depleted CCDs, and have shown through a 
review of the literature that essentially all types of silicon imaging devices can 
exhibit similar effects.  The physics of the observed effects for the various types of 
imagers differ, but in all cases the origin of the low-light variations is the 
resistivity striations that are inherent to the growth of the silicon crystals.  We 
have presented a simple model based on the lateral displacement of charge due to 
lateral volume charge density variations arising from the resistivity striations, 
and have presented experimental evidence in support of the model.  The operation 
of fully depleted CCDs at high vertical electric fields offers a means to minimize 
both effects considered in this paper. 

\acknowledgments

The CCDs used in this work were packaged by John Emes.  We would like to thank 
the anonymous reviewer who made us aware of the prior work on silicon drift
detectors.  This work was supported by the 
Director, Office of Science, of the U.S. Department of Energy under Contract No. 
DE-AC02-05CH11231.  The U.S.  Government retains, and the publisher, by accepting 
the article for publication, acknowledges, that the U.S. Government retains a 
non-exclusive, paid-up, irrevocable, world-wide license to publish or reproduce 
the published form of this manuscript, or allow others to do so, for U.S. 
Government purposes.

 

\end{document}